\documentclass[preprint,aps,pra,showpacs]{revtex4}
\usepackage{tabularx}
\usepackage{dcolumn}
\usepackage{graphics}
\usepackage{bm}
\usepackage[dvips]{graphicx}
\usepackage[dvips]{rotating}
\usepackage[latin1]{inputenc}
\usepackage{dcolumn}
\usepackage{tabularx}
\usepackage{amsmath}
\usepackage{amssymb}
\begin{document}
\draft

\title{Oscillation structures in the spontaneous emission rate
of an atom in a medium with refractive index $n$ between mirrors: a
solvable model}

\author{H. J. Zhao and  M. L. Du} \email{
duml@itp.ac.cn}

\address{Institute of Theoretical Physics, Chinese Academy of Sciences,
P.O.Box 2735, Beijing 100080, China}

\date{\today}

\begin{abstract}
We study the multi-periodic oscillations in the spontaneous emission
rate of an atom in a medium with refractive index $n$ sandwiched
between two parallel mirrors.  The oscillations are not obvious in
the analytical formula for the rate derived based on Fermi's golden
rule but can be extracted using Fourier transforms by varying the
system scale while holding the configuration.  The oscillations are
interpreted as interferences and correspond to various closed-orbits
of the emitted photon going away from and returning to the atom.
This system provides a rare example that the oscillations can be
explicitly derived by following the emitted wave until it returns to
the emitting atom.  We demonstrate the summation over a large number
of closed-orbits converges to the rate formula of golden rule.
\par


\par

\end{abstract}

 \maketitle


\section{Introduction}

The spontaneous emission process of atoms in the presence of
environments is an active research area in recent years.  It is well
known that environments surrounding the atom can greatly modify the
lifetime of atomic state \cite{1}, such modifications are important
in applications.   From a physical point view, the fundamental
problem is to understand the general characteristics in the modified
spontaneous emission rate. Inspired by the similarity between the
spontaneous emission rate and atomic absorption spectra, it has been
shown recently that large scale oscillations in the spontaneous
emission rate of atoms near an dielectric interface and inside a
dielectric slab \cite{2,3} are present and that a theory similar to
closed-orbit theory \cite{4,5,6,7} can be used to understand such
large scale oscillations in the spontaneous emission rate.
Oscillations in the absorption spectra for atoms and negative ions
in external electric and magnetic fields are associated with
electron closed-orbits going away from and returning to the nucleus,
the oscillations in the spontaneous emission rate are associated
with what we call ``photon closed-orbits" going away from and
returning to the emitting atom. Following ``closed-orbit theory",
the oscillations in the spontaneous emission rate are interpreted as
interferences between outgoing emitted electromagnetic wave and
returning electromagnetic wave traveling along various
closed-orbits.

For an atom near an dielectric interface, there is only one
oscillation in the spontaneous emission rate corresponding to one
closed-orbit \cite{3}. The dynamics of this system is quite similar
to the photodetachment of a negative ion in the presence of a static
electric field \cite{8}.  For an atom inside a dielectric slab,
three oscillations were extracted from the numerically calculated
spontaneous emission rates, and these three oscillations were shown
to correspond to three photon closed-orbits \cite{2}.  In this
article we study the oscillations in the spontaneous emission rate
of an atom in a medium with refractive index $n$ between two large
perfect parallel plane mirrors. This system offers several
advantages over an atom inside the dielectric slabs \cite{2}: (1)
the emission rate can be derived analytically; (2) because of the
complete reflections of the mirrors, we are able to extract many
more oscillations from the emission rate and correlate them with
emitted photon closed-orbits; (3) one can follow the emitted wave
and derive a rate formula similar to the closed-orbit theory.

\section{The spontaneous emission rate formula and Fourier transforms}

The schematic diagram of the system is shown in Fig. 1. An atom is
placed between two perfect parallel plane mirrors, and a medium with
refractive index $n$ fills up the space between the mirrors. Denote
the distance between the two mirrors by $d$.  We choose the
coordinate system such that the origin is in the middle of the two
mirrors and the $z$-axis is perpendicular to the mirrors.  The
emitting atom is placed at $z$, its distance from the upper (lower)
mirror is denoted by $d_1$ and $d_2$.  Using Fermi's golden rule and
following the approach developed for the case when the space between
two mirrors is vacuum \cite{9,10,11,12}, the spontaneous emission
rate for this system can be derived. When the transition dipole
moment $\vec{d}_{12}$ is parallel to the mirror planes, the emission
rate formula is
\begin{equation}\label{1}
    W(z)=W_{vac}\frac{3\pi c}{2 d
    \omega_0}\sum_{j=1}^M(1+\frac{j^2\pi^2}{n^2k^2_0d^2})\sin^2(\frac{j\pi
    z}{d}-\frac{j\pi}{2}),
\end{equation}
where $W_{vac}$ is the emission rate in vacuum without mirrors, $M$
is the greatest integer part of $nk_0d/\pi$ , $k_0=\omega_0/c$ is
the wave number of the emitted light in vacuum, $z$ is the
coordinate of the emitting atom. The derivation of formula (1) is
given in Appendix A for completeness.

To investigate the multi-periodic oscillations in the spontaneous
emission rate, we follow the approach for the dielectric slabs
\cite{2} and use $R=(d_2-d_1)/(d_2+d_1)$ to indicate the
configuration of the system. We also use a variable $\alpha$ to
indicate the system size relative to a standard one. We take the
standard distance between the two mirrors as the wavelength of the
emitted photon in vacuum, that is, $d^0=\lambda_0=2\pi/k_0$, then we
have the relationship $z=\alpha R d^0/2$ , $d_1=\alpha(1-R)d^0/2$
and $d_2=\alpha(1+R)d^0/2$. In order to compare with the spontaneous
emission rate in dielectric slabs \cite{2}, we also set the
refractive index $n=1.49$ appropriate for the poly methyl
methacrylate (PMMA) material between the two mirrors. In Fig. 2 we
show the spontaneous emission rate as a function $\alpha$ for
$R=0,1/3$, and $3/5$.  In Fig. 2(a), $R=0$, the emitting atom is in
the middle of the mirrors, the emission rate is like a saw-tooth; In
Fig. 2(b), $R=1/3$, the distance between the emitting atom and the
lower mirror is twice of the distance between the emitting atom and
the upper mirror, some smaller tooth appeared; In Fig. 2(c), the
emitting atom is even closer to the upper mirror, the distance
between the emitting atom and the lower mirror is four times of the
distance between the emitting atom and the upper mirror, finer tooth
appeared. These figures can be compared with the more smooth
emission rate in dielectric slabs \cite{2}.

The oscillations in these figures are not obvious but can be
extracted using modified Fourier transforms defined by
\begin{equation}\label{2}
    \tilde{W}(\gamma)=\int_{\alpha_1}^{\alpha_2}[W(\alpha)-W_{bg}]\alpha
    e^{i\alpha \gamma}d\alpha,
\end{equation}
where $W_{bg}=n W_{vac}$ is the non-oscillating background emission
rate in the medium without mirrors.  In Eq.(2) $W(\alpha)$
represents the dependence of the emission rate as a function of the
scaling variable $\alpha$ . Assuming $\gamma'$ is the first peak
position of the Fourier transform in the range $[\gamma_{min}
,\gamma_{max}]$, the integration range should satisfy
$(\alpha_2-\alpha_1)\geq 2\pi/\gamma'$ and the step size $\Delta
\alpha$ should satisfy $\Delta \alpha\leq
0.1\times2\pi/\gamma_{max}$to ensure numerical accuracy. In our
calculations, we took $\alpha_1=1$, $\alpha_2=16$, $\Delta
\alpha=0.01$. The calculated Fourier transformations of the rates in
Fig.2 as defined in Eq.(2) are presented as solid lines in Fig.3.
The numerical value near each peak is the extracted peak position.
There are five peaks in Fig. 3(a) when $R=0$, seven peaks in Fig.
3(b) when $R=1/3$, and seven peaks in Fig. 3(c) when $R=3/5$. Many
more peaks are present in Fig.3 compared to the dielectric slab
system \cite{2}.

\section{The oscillations and closed-orbits}
We now provide an explanation of the physical meaning for the peaks
and give a quantitative description for the peak positions and peak
heights. Over the last ten years, closed-orbit theory has been used
successfully to understand the oscillations in the absorption
spectra of atoms and negative ions in external fields
\cite{4,5,6,7}. The oscillations in the absorption spectra
correspond to electron closed-orbits, and they are interpreted as
interferences between outgoing electron and returning electron.  The
similarity between the oscillations in the spontaneous emission rate
and the oscillations in the absorption spectra has inspired the
recent studies of the oscillations in the spontaneous emission near
dielectric interfaces \cite{2,3}. The oscillations in the
spontaneous emission rates correspond to closed-orbits of emitted
photon and they are interpreted as interference effects between
emitted photon going away from and returning to the atom following
various closed-orbits. This view can be made more quantitative for
the present system.

Consider the radiation damping of a dipole antenna \cite{13}. The
dipole moment is $\vec{d}e^{-i\omega_0 t}$ , where $\omega_0$ is the
frequency of oscillation. The radiation damping rate is $W_d=P/U$
where $P$ is the average radiation power of the antenna, $U$ is the
energy of the antenna. The radiation damping rate can be written as
\cite{13}
\begin{equation}\label{3}
    W=\frac{\omega_0}{2U}{\rm{Im}}(\vec{d}^*\cdot\vec{E})=\frac{\omega_0}{2U}[{\rm{Im}}(\vec{d}^*\cdot\vec{E}_0)+{\rm{Im}}(\vec{d}^*\cdot\vec{E}_{ret})],
\end{equation}
where we have decomposed the electric field  $\vec{E}$ at the
position of dipole antenna into a direct part $\vec{E}_0$  and a
returning part $\vec{E}_{ret}$.  This step for the electric field is
similar to a step for the electron wave function in the derivation
of closed-orbit theory \cite{4,5,6}.  Using $\vec{r}$ to denote the
vector of a point relative to the dipole position, the direct
electric field is
\begin{equation}\label{4}
   \vec{E}_0=\frac{dk^3}{4\pi \epsilon}\{(\hat{r}\times\hat{d})\times\hat{r}(\frac{1}{kr})
   +[\hat{d}-3\hat{r}(\hat{r}\cdot\hat{d})]\times(\frac{i}{(kr)^2}-\frac{1}{(kr)^3})\}e^{i(kr-\omega_0t)}.
\end{equation}
The returning electric field $\vec{E}_{ret}$ depends on the
environments or the cavity.  If the dipole antenna is at distance
$l$ from the mirror and is parallel to the mirror, $\vec{E}_{ret}$
can be calculated exactly using an imaging method \cite{13}. The
returning field at the position of the antenna is
\begin{equation}\label{5}
    \vec{E}_{ret}=-\frac{dk^3}{4\pi \epsilon}\hat{d}
    (\frac{1}{2kl}+\frac{i}{(2kl)^2}-\frac{1}{(2kl)^3})e^{i(2kl-\omega_0t)},
\end{equation}
where $\hat{d}$ is a unit vector in the dipole direction. The
damping rate for a dipole antenna in a medium with an arbitrary
refractive index $n$ near one mirror and parallel to the mirror
plane can be evaluated as
\begin{equation}\label{6}
    W_d=W_0-\frac{3}{2}W_0[\frac{\sin(2nk_0l)}{2nk_0l}
    +\frac{\cos(2nk_0l)}{(2nk_0l)^2}-\frac{\sin(2nk_0l)}{(2nk_0l)^2}],
\end{equation}
where
$W_0=\frac{2}{3}\frac{d^2k^3}{4\pi\epsilon}\frac{\omega_0}{2U}$ .
Eq.(6) is a generalization of the vacuum case $n =1$ \cite{13}. When
the distance between the dipole antenna and the mirror is greater
than half of a wavelength, the last two terms in the square bracket
are small enough and can be neglected. The semi-classical
approximation for the damping rate of a dipole near one mirror
becomes
\begin{equation}\label{7}
    W_d=W_0-\frac{3}{2}W_0\frac{\sin(S)}{S},
\end{equation}
where $S=2nk_0l$ is the action of the emitted photon going from the
atom to the mirror and back to the atom. This path is a
closed-orbit. For a dipole antenna in a medium with an arbitrary
refractive index $n$ between two mirrors, the semi-classical formula
for the damping rate can be generalized to
\begin{equation}\label{8}
     W_d=W_0-\frac{3}{2}W_0\sum_{j}\frac{1}{S_j}\sin(S_j+\phi_j),
\end{equation}
where $S_j=nk_0L_j$ is the action of the $j$-th emitted photon
closed-orbit, $L_j$ is the corresponding geometric length of the
orbit, $\phi_j=-m_j\pi$ is a phase correction, $m_j$ counts the
number of reflections by the two mirrors. We identify $W_d$ and
$W_0$ in Eq.(8) with the spontaneous rate $W$ and $W_{bg}$ for the
spontaneous emission rate formula in Eq.(1). Eq.(8) is quite similar
to the formula in closed-orbit theory.  The peak positions in the
Fourier transform of Eq.(2) are given by the action $S_j^0$ of
emitted photon closed-orbits with corresponding peak height
proportional  to $g/L_j^0$, where $L_j^0$ is the length of the
closed-orbit, $g$ is a degeneracy factor counting the number of
closed-orbit having the same length.  Both $S_j^0$ and $L_j^0$ are
evaluated for the standard system size corresponding to $\alpha=1$.
We have the scaling relationship $L_j(\alpha)=\alpha L_j^0$ and
$S_j(\alpha)=\alpha S_j^0$.

The above predictions for the peak positions and peak heights in the
Fourier transforms (FT)  are marked as crosses in Fig.3. We fixed
the constant $h$ using one peak height and then calculated other
peak height using $hg/L_j^0$.  For all the peaks in Fig.3, the
predications from Eq.(8) for both the peak positions and the peak
heights agree well with the numerically extracted values from the
rate formula in Eq.(1). For the case $R=0$, there are five peaks.
Each peak corresponds to two closed-orbits. The structures of the
closed-orbits are shown as inserts next to the peaks in Fig. 3(a).
The first peak in Fig. 3(a) corresponds to two closed-orbits going
away from and returning to the emitting atom after being reflected
back by the upper or the lower mirror once.  These two closed-orbits
have the same action $S_{1,2}^2=2\pi n=9.36$; the third (fourth)
closed-orbit leaves from and returns to the emitting atom after
reflected by both mirrors. The action for the third or fourth
closed-orbit is $S_{3,4}^0=4\pi n=18.72$. The fifth (sixth)
closed-orbit is reflected three times by the mirrors. Their actions
are $S^0_{5,6}=6\pi n=28.09$. Similar discussions can be made for
other peaks in Fig.3(a). The double degeneracy for some of the peaks
can be removed when the emitting atom is away from the middle point.
For example, when $R=1/3$, the first peak at $9.35$ in Fig.3(a)
splits to two peaks at $6.23$ and at $12.50$ in Fig.3(b). The peak
at $6.23$ corresponds to the closed-orbit produced by the reflection
of the upper mirror closer to the emitting atom, while the peak at
$12.50$ is produced by the reflection of the lower mirror that is
further away from the emitting atom.  In Table I, we compare the
extracted values and the theoretical predictions for the peak
positions and peak heights in Fig.3.  They agree very well.
\begin{table}
\caption{Peak positions and peak heights extracted from Eq.(1) using
Fourier transform (FT) are compared with the analytic predications
of Eq.(8). }
\begin{tabular}{c c c c c c}
\hline\hline
  ~ & ~ & \multicolumn{2}{c}{Peak position} & \multicolumn{2}{c}{Peak height} \\
  ~ & peak index & FT & Eq. (8) & FT & Eq. (8) \\
   ~ & ~ & ~ & ~ & ~ & ~ \\
  ~ & $1^{st}$ & 9.35 & 9.36 & 3.5987 & 3.5993 \\
  ~ & $2^{nd}$ & 18.72 & 18.72 & 1.8175 & 1.7977\\
  R=0 & $3^{rd}$ & 28.08 & 28.09 & 1.1957 & 1.1984\\
  ~ & $4^{th}$ & 37.45 & 37.45 & 0.90616& 0.8986\\
  ~ & $5^{th}$ & 46.81 & 46.81 & 0.72541& 0.7189\\
  ~ & ~ & ~ & ~ & ~ & ~ \\
  ~ & $1^{st}$  & 6.23  &    6.24  &   2.6839   &  2.7009\\
  ~ & $2^{nd}$  &12.50   &  12.48   &  1.3692  &   1.3461\\
  ~ & $3^{rd}$  &18.72   &  18.72   &  1.8102  &   1.7977\\
R=1/3 & $4^{th}$&24.95  &   24.97  &   0.65365 &0.6744\\
  ~ & $5^{th}$ & 31.22   &  31.21  &   0.55291  &  0.5390\\
  ~ & $6^{th}$ &37.45  &   37.45   &  0.90279  &  0.8986\\
  ~ & $7^{th}$ & 43.67  &   43.69   &  0.37436 &   0.3853\\
 ~ & ~ & ~ & ~ & ~ & ~ \\
    ~ & $1^{st}$ & 3.73&  3.74&   4.5112&     4.5112\\
    ~ & $2^{nd}$ & 14.94&  14.98&     1.1261&    1.1263\\
    ~ & $3^{rd}$ & 18.72&  18.72&     1.7920&    1.7977\\
 R=3/5& $4^{th}$ & 22.50&  22.47&     0.7551&    0.7479\\
    ~ & $5^{th}$ & 33.66&  33.70&     0.50151&   0.4999\\
    ~ & $6^{th}$ & 37.44&  37.45&     0.8917&    0.8989\\
    ~ & $7^{th}$ & 41.22&  41.19&     0.40507&   0.4082\\
\hline\hline
\end{tabular}
\end{table}

In Fig.4 we compare the peak positions extracted from Eq.(1) using
Fourier transform (circles) with the analytic predication (solid
lines) as a function of $R$. 16 peaks can be clearly identified in
the specified range. The numerical and analytic results for the peak
positions are in excellent agreement. The large number of peaks
identified for this system is in contrast with the previous systems
with dielectric interfaces \cite{2,3} where only one or three peaks
were clearly identified. The difference in the number of
identifiable peaks in these systems can be explained by the
difference between a dielectric interface and a mirror in reflecting
emitted photon waves. In deriving Eq.(8),  it is clear that the peak
height is proportional to the wave amplitude when it returns to the
position of the dipole. For a closed-orbit $j$, it is being
reflected $m_j$ times by mirrors or dielectric interfaces. A mirror
reflects better than a dielectric interface, consequently the
returning wave corresponding to the same closed-orbit is stronger in
a mirror system and many more peaks can  be identified.

Eq.(8) is a sum over closed-orbits.  In Fig.5 we show how the sum
over closed-orbits converges to the rate formula in Eq.(1).
Fig.5(a) demonstrates that 4 closed-orbits already gives a rather
accurate representation of the rate that requires a large number of
terms in Eq.(1). When the closed-orbits are increased to 16 as in
Fig.5(b), the oscillations are greatly reduced.  When we use 100
closed-orbits in Eq.(8) as shown in Fig.5(c), the sum over
closed-orbits in Eq.(8) accurately represents the golden rule rate
in Eq.(1).

\section{ Conclusions }

In conclusions, we have studied the oscillations in the spontaneous
emission of an atom in a medium with an arbitrary refractive index
$n$ between two large perfect plane mirrors. We have compared the
golden rule formula in Eq.(1) and the summation formula over
closed-orbits in Eq.(8).  The oscillations in the golden rule
formula can be extracted when the system scale is varied while the
system configuration is fixed.  The extracted peak positions and
peak heights agree well with Eq.(8) based on closed-orbits of
emitted photon going away and returning to the atom.  The
oscillations in the spontaneous emission rate are quite similar to
the oscillations in the absorption spectra described by closed-orbit
theory \cite{4,5,6,7}. For other systems, in principle,  we can
follow the propagation of the electric field along various
closed-orbits until they are back to the emitting atom to calculate
the radiation damping rate using Eq.(3).  This would give us a
formula of the rate in terms of closed-orbits.  Our study suggests
the rate formula based on Fermi's golden rule and the rate formula
involving a sum over closed-orbits provide complementary
perspectives on spontaneous emission process.

\begin{center}
{\bf ACKNOWLEDGMENTS}
\end{center}
\vskip8pt This work was supported by NSFC grant No. 90403028.

\appendix
\section{derivation of formula in Eq.(1)}

The emitting rate is given by Feimi's golden rule \cite{13}. The
transition rate is
\begin{equation}\label{a1}
    W=\frac{2\pi}{\hbar^2}\sum_\alpha|\langle e,0|-\frac{q}{m}\vec{p}\cdot
    \vec{A}(\vec{x}')|q,1_\alpha\rangle|^2\delta(\omega_\alpha-\omega_{eq})
\end{equation}
where $|e,0\rangle$  is the system initial state, in which the atom
is excited and there is no photon in the field; $|q,1_\alpha\rangle$
is the system final state, in which the atom is in final state and
there is one photon with frequency $\omega_\alpha$ in the field; $q$
is the charge of the electron of the atom, and $m$ is the mass of
the electron; $\vec{p}$ is the momentum of the electron, and
$\vec{A}(\vec{x}')$ is the quantized transverse vector potential,
and $\vec{x}'$ is the position of atom; $\omega_{qe}$ is the
frequency $(E_e-E_q)/\hbar$. $\vec{A}(\vec{x})$ can be write as
\begin{equation}\label{a2}
    \vec{A}(\vec{x})=\sum_\alpha
    \sqrt{\frac{\hbar}{2n^2\epsilon_0\omega_\alpha}}[\vec{A}_\alpha(\vec{x})a_\alpha+\vec{A}_\alpha^*(\vec{x})a_\alpha^\dagger]
\end{equation}
where
\begin{equation}\label{a3}
    \bigtriangledown^2\vec{A}_\alpha(\vec{x})+k^2_\alpha\vec{A}_\alpha(\vec{x})=0,~~k_\alpha^2=\frac{\omega_\alpha^2n^2}{c^2},
\end{equation}
\begin{equation}\label{a4}
    \bigtriangledown\vec{A}_\alpha(\vec{x})=0,
\end{equation}
$n$ is the refractive index of the dielectric medium, and the mode
function are chosen to form an orthonormal set:
\begin{equation}\label{a5}
    \int
    d^3x\vec{A}_\alpha^*(\vec{x})\cdot\vec{A}_\beta(\vec{x})=\delta_{\alpha\beta}.
\end{equation}

Using Eqs. (\ref{a1}) and (\ref{a2}) we find
\begin{equation}\label{a6}
 W=\frac{2\pi}{\hbar^2}\sum_\alpha \frac{q^2\hbar}{2m^2\epsilon_0n^2\omega_\alpha}|\vec{p}_{eq}\cdot
    \vec{A}_{\alpha}(\vec{x}')|^2\delta(\omega_\alpha-\omega_{eq})
\end{equation}
where $\vec{p}_{eq}=\langle e|\vec{p}|q\rangle$ . Using
$q\vec{p}_{eq}=-im\omega_{eq}\vec{d}_{eq}$ and Eq. (A6) we get
\begin{equation}\label{a7}
    W= \frac{\pi \omega^2_{eq}}{\hbar\epsilon_0n^2}\sum_\alpha \frac{1}{\omega_\alpha}|\vec{d}_{eq}\cdot
    \vec{A}_{\alpha}(\vec{x}')|^2\delta(\omega_\alpha-\omega_{eq})
\end{equation}
where $\vec{d}_{eq}$ is the electric dipole matrix element.

Consider our boundary conditions: the mirrors are in $z=-d/2$ and
$z=d/2$ , and therefore the mode function must vanish at this two
planes. According to Ref. \cite{10} we write the mode functions as
\begin{equation}\label{a8}
    A_{\vec{k}_1}(\vec{x})=(\frac{2}{V})^{1/2}(\hat{k}_\|\times
    \hat{z})\sin k_3(z-\frac{d}{2})e^{i\vec{k}_{\|}\cdot\vec{x}}
\end{equation}
\begin{equation}\label{a9}
    A_{\vec{k}_2}(\vec{x})=\frac{1}{k}(\frac{2}{V})^{1/2}[k_\|
    \hat{z}\cos k_3(z-\frac{d}{2})-ik_3\hat{k}_\|
   \sin k_3(z-\frac{d}{2})]e^{i\vec{k}_{\|}\cdot\vec{x}}
\end{equation}
where $\vec{k}_{\|}$ is the parallel component of $\vec{k}$ , and
$\vec{k}_3=j\frac{\pi}{d},(j=1.2.\cdots)$ . We consider
$\vec{d}_{eq}$ is parallel the mirror and set $\phi$ as the angle
between $\vec{k}_{\|}$ and $\vec{d}_{eq}$. Then we get
\begin{eqnarray}\label{a10}
 |\vec{A}_{\alpha}(\vec{x})\cdot\vec{d}_{eq}
    |^2 &=& \frac{2}{V}[|(\hat{k}_\|\times
    \hat{z})\cdot\vec{d}_{eq}|^2\sin^2 k_3(z-\frac{d}{2})+
    \frac{k_3^2}{k_\alpha^2}|\hat{k}_\|\cdot\vec{d}_{eq}|^2\sin^2
    k_3(z-\frac{d}{2})]\\\label{a11}
  ~ &=& \frac{2|\vec{d}_{eq}|^2}{V}[(\sin^2\phi+\frac{k_3^2}{k_\alpha^2}\cos^2\phi)\sin^2k_3(z-\frac{d}{2})]
\end{eqnarray}
Using Eq. (\ref{a7}) and (\ref{a11}), we get
\begin{eqnarray}
  \nonumber W &=& \frac{\pi \omega^2_{eq}}{\hbar\epsilon_0n^2}\sum_{k_3}\frac{L^2}{4\pi^2}
  \int_0^\infty d k_{\|}k_{\|}\int_0^{2\pi}d\phi\frac{2|\vec{d}_{eq}|^2}{V}
  [(\sin^2\phi+\frac{k_3^2}{k_\alpha^2}\cos^2\phi)\\
  ~ &~&\times\sin^2k_3(z-\frac{d}{2})]
  \frac{1}{\omega_\alpha}\delta(\omega_\alpha-\omega_{eq})\\
   \nonumber~ &=& \frac{\pi \omega^2_{eq}}{\hbar\epsilon_0n^2}\frac{L^2}{4\pi^2}
   \frac{2|\vec{d}_{eq}|^2}{V}\sum_{k_3}\int_0^\infty d k_{\|}k_{\|}
   \pi(1+\frac{k_3^2}{k^2_\alpha})\\
  ~ &~&\times\sin^2k_3(z-\frac{d}{2})\frac{n}{ck}
   \frac{n^2k_0}{ck_{\|}}\delta(k_\|-\sqrt{n^2k_0^2-k_3^2})\\
  ~ &=& \frac{\pi \omega^2_{eq}}{\hbar\epsilon_0n^2}\frac{L^2}{4\pi^2}
   \frac{2|\vec{d}_{eq}|^2}{V}\sum_{k_3}\pi(1+\frac{k_3^2}{n^2k_0^2-k_3^2+k_3^2})\sin^2k_3(z-\frac{d}{2})
   \frac{n}{ck}\frac{n^2k_0}{c}\\
  ~ &=& \frac{\pi \omega^2_{eq}}{\hbar\epsilon_0n^2}\frac{L^2}{4\pi^2}
   \frac{2|\vec{d}_{eq}|^2}{V}\frac{n}{ck}\frac{n^2k_0}{c}\pi\sum_{j}
   (1+\frac{j^2\pi^2}{d^2n^2k_0^2})\sin^2(\frac{j\pi z}{d}-\frac{j\pi}{2}) \\
  ~ &=& W_{vac}\frac{3\pi}{2k_0d}\sum_{j}^{M}(1+\frac{j^2\pi^2}{d^2n^2k_0^2})\sin^2(\frac{j\pi z}{d}-\frac{j\pi}{2})
\end{eqnarray}
where
$W_{vac}=\frac{\omega_{eq}^3|\vec{d}_{eq}|^2}{3\hbar\epsilon_0\pi
c^3}$ is the emission rate in vacuum.


\newpage
\begin{figure}
\includegraphics[scale=.80,angle=-0]{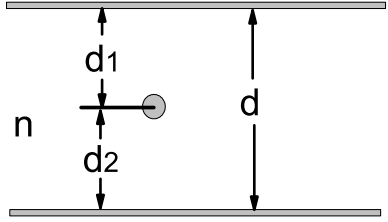}
\caption{Schematic diagram for emitting atom in a medium with an
arbitrary refractive index $n$  between two perfect parallel plane
mirrors. }
\end{figure}

\begin{figure}
\includegraphics[scale=.80,angle=-0]{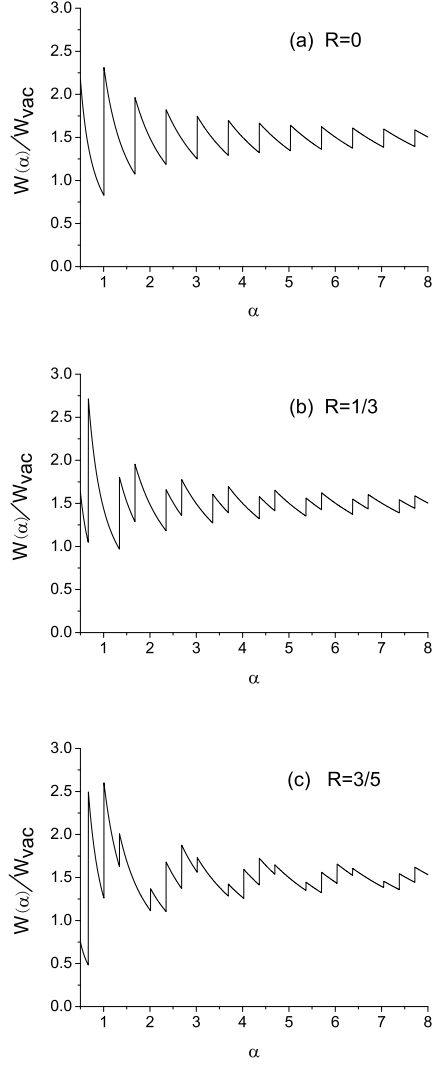}
\caption{ Spontaneous emission rate (in unit of rate in vacuum) as a
function of $\alpha=d/d^0$ , where $d^0$ is the emitted photon
wavelength in vacuum. The three cases in (a), (b), and (c)
correspond to $R=(d_2-d_1)/(d_2+d_1) =0$, $1/3$, and $3/5$,
respectively. }
\end{figure}

\begin{figure}
\includegraphics[scale=.80,angle=-0]{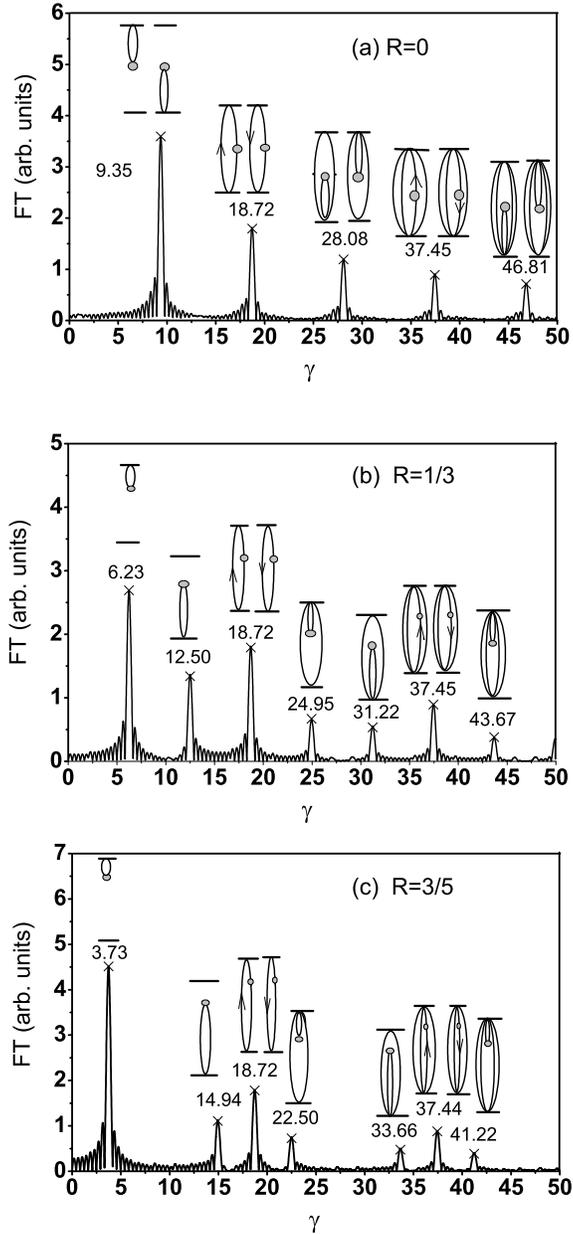}
\caption{ Fourier transforms of the spontaneous emission rate in
Fig. 2 (solid lines) are compared with the analytical results
(crosses) based on closed-orbits. Closed-orbits corresponding to the
peaks are shown schematically in the inserts.  }
\end{figure}

\begin{figure}
\includegraphics[scale=.80,angle=-0]{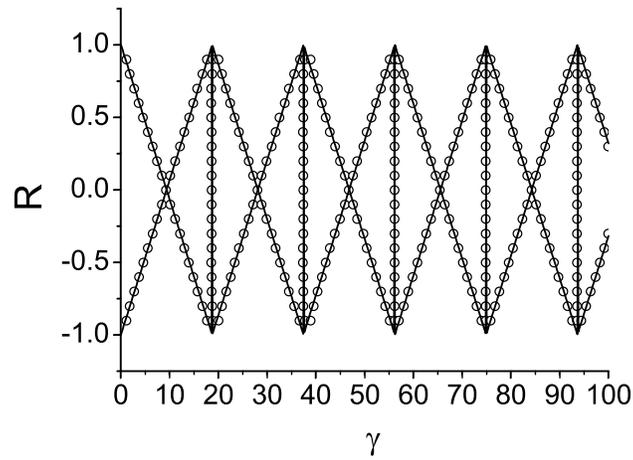}
\caption{ Peak positions extracted from Eq.(1) using Fourier
transforms (circles) are compared with the predictions of
closed-orbits in Eq.(8) as a function of $R$. 16 visible peaks can
be clearly identified for most of $R$ in this range.  }
\end{figure}

\begin{figure}
\includegraphics[scale=.80,angle=-0]{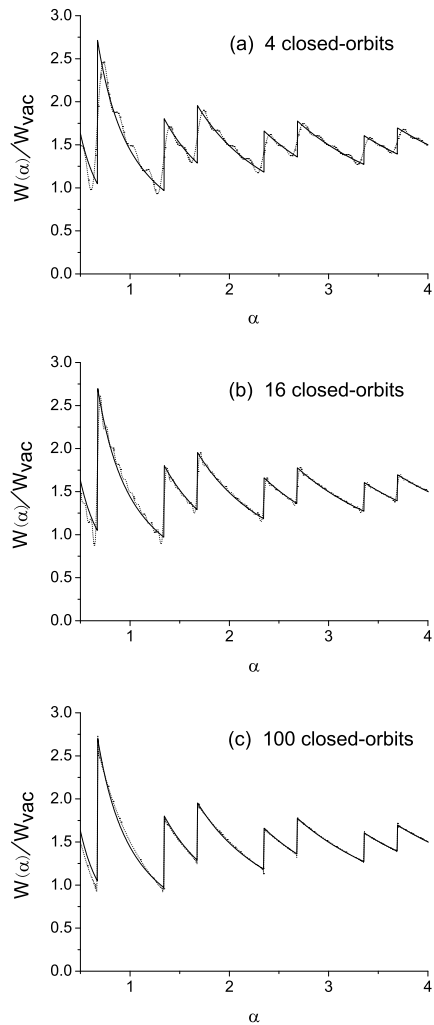}
\caption{ The spontaneous emission rate in Eq.(1) (solid lines) and
the representation using a finite number of closed-orbits in Eq.(8)
(dotted lines).  (a) $4$ closed-orbits; (b) $16$ closed-orbits; (c)
$100$ closed-orbits.  }
\end{figure}


\end{document}